\documentclass{iaus}
\usepackage{graphicx}
\usepackage{natbib}

\title[TeV gamma-rays from LS 5039] %% give here short title %%
{Modeling TeV gamma-rays from LS 5039: \\ An active OB star at the extreme}

\author[Stan Owocki, Atsuo Okaazki \& Gustavo Romero]   %% give here short author list %%
{Stan Owocki$^1$,
Atsuo Okazaki$^2$,
and Gustavo Romero$^3$}

\affiliation{
$^1$
Bartol Research Institute, 
Department of Physics \& Astronomy, 
University of Delaware \\
Newark, DE 19716, USA 
\\ email: {\tt owocki@udel.edu} \\[\affilskip]
$^2$
Faculty of Engineering, 
Hokkai-Gakuen University \\
Toyohira-ku, 
Sapporo 062-8605, Japan 
\\email: {\tt okazaki@elsa.hokkai-s-u.ac.jp}  \\[\affilskip]
$^{3}$
Facultad de Ciencias Astron\'omicas y Geof\'{\i}sicas, 
Universidad Nacional de La Plata \\
Paseo del Bosque, 
1900 La Plata, 
Argentina
\\email: {\tt romero@fcaglp.unlp.edu.ar}
}

\pubyear{2010}
\volume{xxx}  %% insert here IAU Symposium No.
\pagerange{xxx--xxx}
\jname{Active OB Stars: Structure, Evolution, Mass Loss \& Critical
Limits}
\editors{C. Neiner, G. Wade, G. Meynet \& G. Peters, eds.}

\begin{document}

\maketitle

\begin{abstract}
Perhaps the most extreme examples of ``Active OB stars" are the subset
of high-mass X-ray binaries -- consisting of an OB star plus compact
companion -- that have recently been observed 
by {\em Fermi} and ground-based Cerenkov telescopes like {\em HESS}
to be sources of very
high energy (VHE; up to 30 TeV!) $\gamma$-rays.  
This paper focuses on the prominent $\gamma$-ray source, LS5039, 
which consists of a massive O6.5V star in a 3.9-day-period, 
mildly elliptical ($e \approx 0.24$)
orbit with its companion, assumed here to be a black-hole or
unmagnetized neutron star.  
Using 3-D SPH simulations of the Bondi-Hoyle accretion of the O-star 
wind onto the companion, we find that the orbital phase variation of 
the accretion follows very closely the simple Bondi-Hoyle-Lyttleton (BHL) rate
for the local radius and wind speed.  
Moreover, a simple model, wherein intrinsic emission of
$\gamma$-rays is assumed to track this accretion rate, reproduces
quite well {\em Fermi} observations of the phase variation of
$\gamma$-rays in the energy range 0.1-10 GeV.
However for the VHE (0.1-30 TeV) radiation observed by the {\em HESS} Cerenkov
telescope, it is important to account also for photon-photon interactions between
the $\gamma$-rays and the stellar optical/UV radiation, which effectively attenuates
much of the strong emission near periastron. 
When this is included, 
% we find that this simple BHL accretion model 
% quite naturally fits both the {\em Fermi} and {\em HESS} light curves, 
we find that this simple BHL accretion model also quite
naturally fits the {\em HESS} light curve, 
thus making it a strong alternative to the pulsar-wind-shock models commonly
invoked to explain such VHE $\gamma$-ray emission in massive-star binaries.

\end{abstract}

\firstsection % if your document starts with a section,

\section{Introduction}

Among the most active OB stars are those in high-mass X-ray binary (HMXB)
systems, in which 
interaction of the massive-star wind
with a close compact companion 
-- either a neutron star or black hole -- 
produces hard ($>10$~keV) X-ray emission with characteristic,
regular modulation over the orbital period.
In recent years a small subset 
of such HMXB's have been found
also to be gamma-ray sources, with energies up to $\sim$10~GeV
observed by orbiting gamma-ray observatories like {\em Fermi},
and very-high-energies (VHE) of over a TeV ($10^{12}$~eV) seen by ground-based
Cerenkov telescopes like {\em HESS}, {\em Veritas}, and {\em Magic}.
For the one case, known as B\,1259-63, in which oberved radio pulses show the
companion to be a pulsar, the $\gamma-$ray emission seems 
best explained by a {\em Pulsar-Wind-Shock} (PWS) model,
wherein the interaction of the relativistic pulsar wind with the dense
wind of the massive-star produces strong shocks that accelerate
electrons to very high energies, with inverse-Compton scattering of the stellar 
light by these high-energy electrons then producing VHE $\gamma$-rays.
For the other gamma-ray binaries, pulses have not been detected, 
and the nature of the companion, and the applicability of the PWS model, 
are less clear. 

An alternative{\em MicroQuasar} (MQ) model instead posits that accretion of
circumstellar and/or wind material from the massive star onto the companion
-- assumed now to be either a black hole or a weakly magnetized
neutron star --
powers a jet of either relativistic protons, which interact with
stellar wind protons to produce pions that quickly decay into
$\gamma$-rays, or relativistic pair plasma, which inverse Compton
scatters stellar photons to $\gamma$-rays.
% powers a jet of relativistic protons, which interact with stellar wind
% protons to produce pions that quickly decay into $\gamma$-rays.
A poster paper by Okazaki et al.\ in these proceedings uses Smoothed
Particle Hydrodyanmics (SPH) simulations to examine both the PWS and
MQ models, applying them respectively for two systems,  
B~1259-63 and LSI~+61~303,
in which the massive star is a Be star.
In such Be binary systems, the compact companion can interact with the 
Be star's low-density polar wind, and its dense equatorial decretion disk,
and so modeling both types of interaction is key to
to understanding their high-energy emission.

In the paper here, we apply the same SPH code to a MQ model for
the TeV-binary LS~5039, for which the (non-pulsing) compact companion
is in a moderately eccentric ($e \sim 0.24$),  $3.9$-day orbit
around a massive, non-Be primary star of spectral type O6.5V.
(See Table~1 for full parameters.)
This builds on our previous study \citep{ORO08} to account now for
a fixed wind acceleration, instead of just assuming a constant wind speed.
A key result is that the orbital variation of the 3-D SPH accretion rate of 
the stellar wind flow onto the compact companion follows very closely 
the analytic Bondi-Hoyle-Lyttleton (BHL) rate that depends on the
wind speed and orbital separation at each phase.
(See right panel of Figure~\ref{fig:mdot}.)
Assuming the mass accretion translates promptly into jet power and
thus $\gamma$-ray emission, we then apply this result to derive
predicted light curves at both the GeV energies oberved by {\em Fermi} and 
the TeV energies observed by {\em HESS}.
For the former, we find that assuming GeV $\gamma$-ray emission tracks
closely the BHL accretion rate gives directly a quite good fit to the 
{\em Fermi} lightcurve 
\citep[see left panel of Figure \ref{fig:sim-vs-fermi-hess} below]{abdoetal09}.
But for the latter case one must also account for the attenuation of
emitted TeV $\gamma$-rays by photon-photon interaction with the
optical and UV radiation of the massive star.
When this is included, then the predicted TeV lightcurve also closely
matches the  {\em HESS} observations 
\citep[see right panel of Figure \ref{fig:sim-vs-fermi-hess} below]{aha06}.

\begin{table}[!ht]
\caption{Model parameters for LS~5039}
\centerline{
\begin{tabular}{lcc}
\hline
 & Primary & Secondary \\
\hline
\hline
Spectral type & O6.5V & Compact \\
Mass ($M_{\odot}$) & 22.9$^{\rm a}$ & 3.7$^{\rm a}$ \\
Radius & $9.3 R_{\odot}$$^{\rm a}$ ($=0.31a$) & $2.5 \times 10^{-3}a$ \\
Effective temperature $T_\mathrm{eff}$ (K) & 39,000$^{\rm a}$ & -- \\
Wind terminal speed  $V_{\infty}$ (${\rm km}\,{\rm s}^{-1}$) &
   2440 & -- \\
Mass loss rate $\dot{M}_{*}$ ($M_{\odot}\,{\rm yr}^{-1}$) & 
   $5 \times 10^{-7}$\,\,$^{\rm a}$ &  -- \\
Orbital period $P_{\rm orb}$ (d) & \multicolumn{2}{c}{3.9060$^{\rm a}$} \\
Orbital eccentricity $e$ & \multicolumn{2}{c}{0.24$^{\rm b}$} \\
Semi-major axis $a$ (cm) & \multicolumn{2}{c}{$2.17 \times 10^{12}$}\\
\hline
\multicolumn{3}{l}{$^{\rm a}$ \citet{cas05} $^{\rm b}$ \citet{Szal10}}.
\end{tabular}}
\label{tbl:params}
\end{table}

\section{Bondi-Hoyle-Lyttleton (BHL) Accretion}

\begin{figure}[!ht]
\centerline{
\includegraphics*[width=1\textwidth]{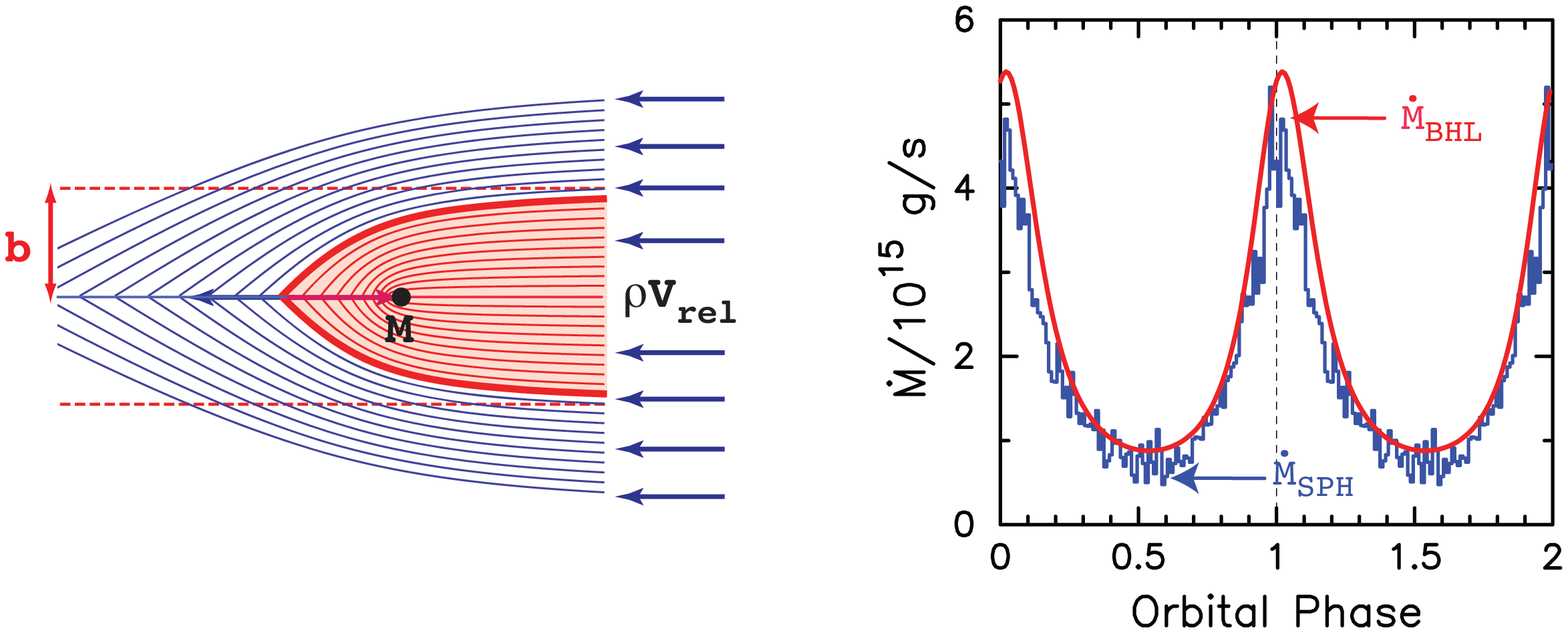}
}
\caption{
{\em Left:} Illustration of flow streams for simple planar form of
Bondi-Hoyle accretion.
The gravitational attraction of mass $M$ focuses an inflow
with relative speed $V_{rel}$ onto the horizontal
symmetry axis.
This causes material impacting within a 
BHL radius $b=2GM/V_{rel}^{2}$ of the axis to be accreted onto $M$,
as outlined here by the bold (red) curve
that bounds the lightly shaded region representing accreting material.
{\em Right:}
Orbital phase variation of the mass-accretion rate from
3-D SPH simulations (blue), compared to the analytic BHL formula from
eqn.\ (\ref{eq:mdotbin}) (red) 
assuming the parameters given in Table~\ref{tbl:params}.
}
\label{fig:BHL-focusing}
\label{fig:mdot}
\end{figure}

To provide a physical context for modeling accretion in LS~5039, 
it is helpful first to review briefly the basic scalings for accretion of an
incoming flow onto a gravitating body, 
as analyzed in  pioneering studies by Bondi, Hoyle and Lyttleton (BHL)
\citep[see][for references and a modern review]{ed04}.
The left panel of Figure~\ref{fig:BHL-focusing} illustrates the basic model.
An initially laminar flow with density $\rho$ 
and relative speed $V_{rel}$ is focussed by the gravity of a mass $M$
onto the downstream  side of the flow symmetry axis,
whereupon the velocity component normal to the axis is cancelled by
collisions among the streams from different azimuths.
Material with initial impact parameter equal to a critical value 
$b \equiv 2GM/V_{rel}^{2}$
-- now typically dubbed the BHL radius, 
and highlighted by the (red) heavy curve in Figure~\ref{fig:BHL-focusing} -- 
arrives on the axis a distance $b$ from the mass, 
with a parallel flow energy $V_{rel}^{2}/2$ just equal to the gravitational 
binding energy $GM/b$.
Since all material with an impact $<b$ (i.e. within the shaded area) should 
thus have negative total
axial energy, it should be eventually accreted onto the central object.
This leads to a simple BHL formula for the expected mass accretion rate,
\begin{equation}
{\dot M}_{BHL} = \rho V_{rel} \pi b^{2} =  
\frac{4 \pi \rho G^{2} M^{2}}{V_{rel}^{3}}
\, .
\label{eq:mdotbhl}
\end{equation}

In a binary system like LS~5039 there arise additional effects from orbital
motion and the associated coriolis terms.
But if we ignore these and other complexities, we can use eqn.\ (\ref{eq:mdotbhl}) to
estimate how the accretion rate should change as a result of the
changes in binary separation $r$ over the system's elliptical orbit,
\begin{equation}
{\dot M}_{BHL} = {\dot M}_{w} 
\frac{ G^{2} M^{2}}{r^{2} V_{w} V_{rel}^{3}}
\, ,
\label{eq:mdotbin}
\end{equation}
where ${\dot M}_{w} \equiv 4 \pi r^{2} \rho (r) V_{w} (r)$ is the wind 
mass loss rate, and $V_{rel}(r)$ is the local magnitude of the 
relative wind and orbital velocity vectors. 
We assume here that the stellar wind follows a standard `beta-type' velocity law,
$V_{w} (r) = V_{\infty} (1-R_{\ast}/r)^{\beta}$,
where $R_{\ast}$ is the O-star radius, 
$V_{\infty}$ is the wind terminal speed, and we adopt here a 
standard velocity power index $\beta=1$.

For the LS~5039 system parameters given in Table~\ref{tbl:params},
the smooth (red) curve in the right panel of Figure~\ref{fig:mdot}
plots the orbital phase variation of this analytically predicted 
BHL-wind-accretion rate.
A principal result of this paper is that this is in remarkably close
agreement with the jagged (blue) curve, 
representing the corresponding accretion rate from  full, 3-D SPH simulations, 
the details of which we discuss next.

\begin{figure}[!ht]
\centerline{
\includegraphics*[width=1\textwidth]{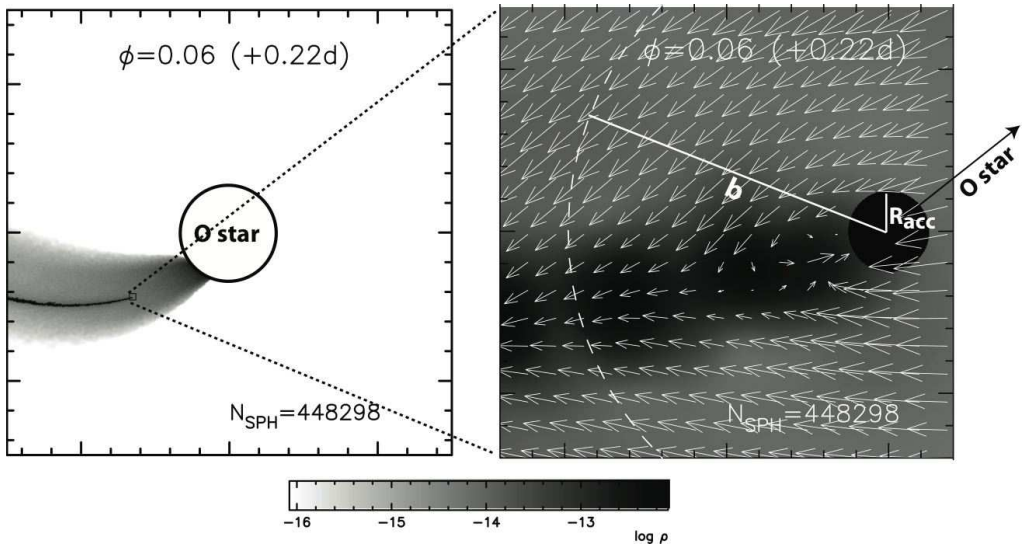}
}
\caption{
Snapshots near periastron in SPH simulation of LS~5039, with the
grayscale showing 
log of the density (in $g/cm^{3}$),
and the arrows denoting flow direction and speed.
The left panel gives an overall view of the
O-star wind stream in the direction of the orbiting companion, which
focuses the impingent wind flow into a narrow, dense, downstream wake.
The right panel zooms in on this wake at
a much smaller scale, within a BHL accretion radius $b$
(marked by the dashed white circle arc) from the companion
(represented by the black sphere of assumed accretion radius
$R_{acc}$).
The white arrows show how dense material close to the
companion and along the  gravitational focus axis forms an accretion stream 
onto the companion, 
much as in the simple laminar flow picture illustrated in the left panel of
Figure~\ref{fig:BHL-focusing}.
}
\label{fig:orbit+acc}
\end{figure}

\section{SPH Simulations}

The SPH code used here is based on a version 
originally developed by 
% \citet{ben90a}, 
\citet{ben90b} and \citet{bat95}, 
with recent extensions to model interacting binaries by
\citet{oka08}.
% and \citet{rom07}. 
It uses a variable smoothing length and integrates the SPH equations with
the standard cubic-spline kernel using individual time steps for each particle. 
The artificial viscosity parameters are set to standard values of
$\alpha_{\rm SPH}=1$ and $\beta_{\rm SPH}=2$. 
In the implementation here, the O-star wind is modeled by
an ensemble of isothermal gas particles of negligible mass, 
while the compact object is represented 
by a sink particle of mass $M$ and  radius $R_{acc} = 5 \times
10^{9}$~cm, i.e. much larger than the $\sim 10^{6}$~cm radius of a
compact object, but still a factor 10 smaller than the minimum (periastron) 
value 
of the BHL accretion radius $b$;
if SPH particles fall within this accretion sphere, 
they are removed from the simulation.
The O star's mass 
exerts a gravitational pull on the 
binary companion, and the net force of radiative driving vs.\
gravity of the O-star leads to the outward wind acceleration characterized 
by the assumed beta=1 velocity law.
In addition, the individual SPH particles feel the gravitational
attraction from the compact companion.
Finally, to optimize the resolution and computational efficiency of our simulations, 
the wind particles are ejected only in a narrow range of azimuthal and vertical angles 
toward the companion;
figure~1 of \citet{ORO08} shows that this gives quite similar accretion
flow structure to what is obtained in a full, spherically symmetric
model, with however many fewer required particles.

For the system parameters listed
in Table~\ref{tbl:params},
Figure~\ref{fig:orbit+acc} illustrates the nature of the accretion
in these SPH simulations, 
using a time snapshot near periastron, with phase $\phi=0.06$.  
The left panel shows the overall, orbitally deflected wind
stream that flows radially away from the O-star toward the companion, 
along with the narrow, dense, gravitationally focussed wind-stream in its wake.
The right panel zooms in on this wake on a scale within a BHL accretion radius
of the companion, as denoted by the dashed white circle arc.
The white arrows show that, within the portions of this dense focal stream
wake nearest the companion, the flow becomes directed toward the 
black sphere with assumed accretion radius $R_{acc}$.

Averaged over some detailed variations, the overall process of
accretion in this fully 3-D SPH simulation is thus remarkably similar to the
simple laminar flow picture illustrated in the left panel of
Figure~\ref{fig:BHL-focusing}.
The right panel of Figure~\ref{fig:mdot} shows moreover that the SPH accretion 
rate
-- averaged over a phase interval 0.01 to smooth over rapid fluctuations --
also agrees remarkably well with the simple BHL rate 
given by eqn.\ (\ref{eq:mdotbin}).

\begin{figure}[!ht]
\centerline{
\includegraphics*[width=1\textwidth]{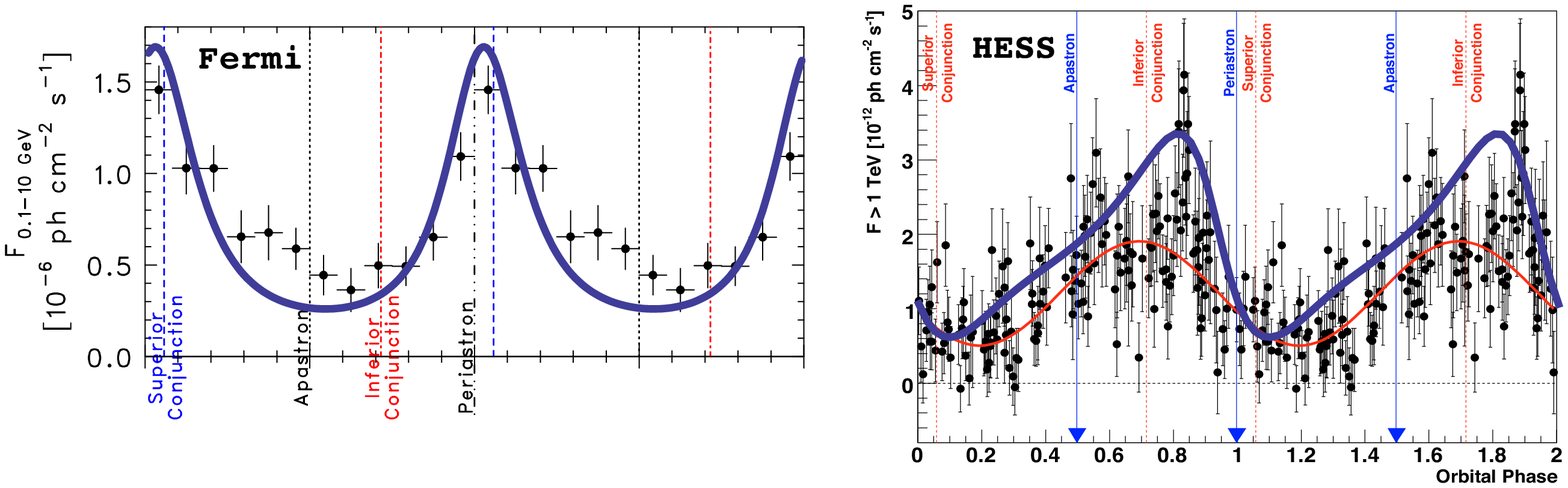}
}
\caption{
LS~5039 light curves at energies 0.1-10~GeV observed by {\em Fermi} 
\citep[left]{abdoetal09}, and 
at energies above 1~TeV observed by {\em HESS} \citep[right]{aha06}, 
plotted vs. orbital phase, and compared against the model
results based on BHL accretion, shown as heavy solid (blue) curves. 
(The lighter (red) curve represents the best-fit sinusoid for the {\em
HESS} data.)
}
\label{fig:sim-vs-fermi-hess}
\end{figure}

\section{Accretion-powered $\gamma$-ray emission}

In a microquasar model, the accretion onto the compact companion
powers a high-energy jet that produces $\gamma$-rays.
Since the jet acceleration occurs on a scale of a few tens of compact
companion radii, it seems reasonable to postulate that the $\gamma$-ray
emission could promptly track the accretion rate.
The left panel of Figure~\ref{fig:sim-vs-fermi-hess} compares the {\em
Fermi} lightcurve for LS~5039 at energies 0.1-10~GeV with a simple
emission model that scales directly with the BHL accretion rate.
The model fits the data quite well, reproducing both the roughly
factor five variation range of observed $\gamma$-rays, as well as the 
observed flux peak at a phase very near periastron\footnote{
If instead of an eccentricity $e=0.24$, we use the larger
value $e=0.35$ suggested by \citet{cas05}, the amplitude of orbital
variation in mass accretion rate then becomes stronger than this
factor five $\gamma$-ray variation observed by {\em Fermi}.
}.

By contrast, the right panel of  Figure~\ref{fig:sim-vs-fermi-hess}
shows that the TeV $\gamma$-rays observed from LS~5039 by {\em HESS} peak well
before periastron, toward the phase associated with inferior
conjunction, with the flux minimum occuring just after periastron, 
near the phase for {\em superior} conjunction.
These phase shifts in the peak and minimum can be explained by
accounting for the attenuation of the $\gamma$-ray rays through
photon-photon interaction with the radiation from the 
O-star.
For TeV $\gamma$-rays, the geometric mean of the energies of the
$\gamma$-rays and stellar UV/optical photons exceeds twice the
rest-mass-energy of electrons, thus allowing photon-photon production 
of electron/positron pairs.
The thick (blue) curve in Figure~\ref{fig:sim-vs-fermi-hess} shows that 
a simple model combining a BHL-rate emission with such photon-photon 
attenuation does indeed match the {\em HESS} data very well.
In contrast, for GeV energy $\gamma$-rays the geometric mean with
stellar UV/optical light generally falls below the electron
mass-energy, and so the {\em Fermi} lightcurve is unaffected by such
attentuation.

Since the cross section for photon-photon interaction is highest near the
threshhold of about 0.1~TeV, a pure attenuation model predicts that
the {\em HESS} spectrum should be hardest during the superior conjunction 
phase of minimum flux and maximum attenuation.
However, plots of the photon index for TeV observations by {\em HESS} 
show just the opposite, with the minimum flux corresponding to the
softest spectrum.
We are currently investigating whether this can be explained by
accounting for the progressive softening of $\gamma$-rays 
during a cascade of absorption and reemission associated with pair
production and anniliation.

\begin{discussion}

\discuss{Doug Gies}{In Bondi-Hoyle accretion, the gas falls towards the black hole
and presumably forms a small accretion disk that powers the
relativistic jets. ÊIn your model, how long does the gas reside
in the disk before being launched into the jets?}

\discuss{Stan Owocki}{Our modeling thus far effectively assumes this
residence time is short, but this requires further investigation
through a more detailed model of the accretion powering of the jet.}

\discuss{Guillaume Dubus}{
I know of no accretion model showing that gamma ray emission from accretion can promptly follow
the orbital variation of the BHL accretion rate. Ê
Also, how can the MQ model explain the low fraction of gamma-ray sources among X-ray
binaries? 
The PWS model explains this through the fact they exist for only a short time in a 
state that can give the required relativistic wind.
I thus believe such PWS models are much favored over an MQ model.
}

\discuss{Stan Owocki}{
Our co-author on this paper,
Gustavo Romero, who could not attend this meeting, has voiced to us
similarly strong arguments in favor of the MQ model over the PWS model
for LS5039.  I expect there will be a vigorous debate on these issues
at the upcoming gamma-ray meeting in Heidelberg this December.
}

\end{discussion}

\end{document}